\documentclass[showpacs  ,secnumarabic, nobibnotes,  nofootinbib, aps,
twocolumn, prd]{revtex4}%
\usepackage{amsmath}
\usepackage{amssymb}
\usepackage{bm}
\usepackage{graphicx}
\usepackage{amsfonts}%
\providecommand{\U}[1]{\protect\rule{.1in}{.1in}}

\begin{document}
\title{Spin-orbital composition in relativistic many-fermion systems}
\author{Petr Zavada}
\affiliation{Institute of Physics AS CR, Na Slovance 2, CZ-182 21 Prague 8, Czech Republic}

\begin{abstract}
The interplay of spins and orbital angular moments of the fermions play an
important role for the structure of the many-fermion systems like atoms,
nuclei, nucleons (baryons) or mesons. We start our study from the one-fermion
eigenstates of angular momentum represented by the spinor spherical harmonics.
Afterwards we study the properties of many-fermion states resulting from a
multiple angular momentum composition of the one-fermion states, giving the
total angular momentum $J=\left\langle L\right\rangle +\left\langle
S\right\rangle $, which is identified with the spin of the composite particle.
We demonstrate how the composition rules affect the relativistic interplay
between the sums of the spins $\left\langle S\right\rangle $ and orbital
angular moments $\left\langle L\right\rangle $ of the constituents, which
collectively generate the spin of composite particle. It is suggested that in
a relativistic case, when the masses of the constituent fermions are much less
than their energy (in the rest frame of the composite particle), then the spin
of the composite particle is dominated by the orbital angular moments
$\left\langle L\right\rangle $ of the constituents, while $\left\vert
\left\langle \mathbb{S}\right\rangle \right\vert \leq$ $J/3$. A special
attention is paid to the case $J=1/2$ that is related to the spin of proton
generated by the composition of spins and orbital angular moments of the quarks.

\end{abstract}

\maketitle

\section{Eigenstates of angular momentum}

\label{eigenstatesAM}The solutions of free Dirac equation represented by
eigenstates of the total angular momentum (AM) with quantum numbers $j,j_{z}$
are the spinor spherical harmonics \cite{bdtm,lali, bie}, which in the
\textit{momentum representation} reads%
\begin{equation}
\left\vert j,j_{z}\right\rangle =\Phi_{jl_{p}j_{z}}\left(  \mathbf{\omega
}\right)  =\frac{1}{\sqrt{2\epsilon}}\left(
\begin{array}
[c]{c}%
\sqrt{\epsilon+m}\Omega_{jl_{p}j_{z}}\left(  \mathbf{\omega}\right)  \\
-\sqrt{\epsilon-m}\Omega_{j\lambda_{p}j_{z}}\left(  \mathbf{\omega}\right)
\end{array}
\right)  ,\label{rs1}%
\end{equation}
where $\mathbf{\omega}$ represents the polar and azimuthal angles
($\theta,\varphi$) of the momentum $\mathbf{p}$ with respect\ to the
quantization axis $z,$ $l_{p}=j\pm1/2,\ \lambda_{p}=2j-l_{p}$ ($l_{p}$ defines
the parity), energy $\epsilon=\sqrt{\mathbf{p}^{2}+m^{2}}$, and
\begin{align}
\Omega_{jl_{p}j_{z}}\left(  \mathbf{\omega}\right)   &  =\left(
\begin{array}
[c]{c}%
\sqrt{\frac{j+j_{z}}{2j}}Y_{l_{p},j_{z}-1/2}\left(  \mathbf{\omega}\right)  \\
\sqrt{\frac{j-j_{z}}{2j}}Y_{l_{p},j_{z}+1/2}\left(  \mathbf{\omega}\right)
\end{array}
\right)  ;\ l_{p}=j-\frac{1}{2},\label{rs1e}\\
\Omega_{jl_{p}j_{z}}\left(  \mathbf{\omega}\right)   &  =\left(
\begin{array}
[c]{c}%
-\sqrt{\frac{j-j_{z}+1}{2j+2}}Y_{l_{p},j_{z}-1/2}\left(  \mathbf{\omega
}\right)  \\
\sqrt{\frac{j+j_{z}+1}{2j+2}}Y_{l_{p},j_{z}+1/2}\left(  \mathbf{\omega
}\right)
\end{array}
\right)  ;\ l_{p}=j+\frac{1}{2}.\nonumber
\end{align}
In a relativistic case the quantum numbers of spin and orbital angular
momentum (OAM) are not conserved separately, but only the total AM $j$ and its
projection $j_{z}=s_{z}+l_{z}$ can be conserved. The complete wave function
reads%
\begin{equation}
\Psi_{jl_{p}j_{z}}\left(  \epsilon,\mathbf{\omega}\right)  =\phi_{j}\left(
\epsilon\right)  \Phi_{jl_{p}j_{z}}\left(  \mathbf{\omega}\right)
.\label{rs1a}%
\end{equation}
The function $\phi_{j}\left(  \epsilon\right)  $\ or its equivalent
representation (\ref{rs1f}) is the amplitude of probability that the fermion
has energy \ $\epsilon$. In fact the main results in this note depend only on
the probability distribution $a_{j}^{\ast}\left(  \epsilon\right)
a_{j}\left(  \epsilon\right)  $ via the parameters (\ref{rs8b}). The spinors
(\ref{rs1}) are normalized as%
\begin{equation}
\int\Phi_{j^{\prime}l_{p}^{\prime}j_{z}^{\prime}}^{+}\left(  \mathbf{\omega
}\right)  \Phi_{jl_{p}j_{z}}\left(  \mathbf{\omega}\right)  d\mathbf{\omega
}=\delta_{j^{\prime}j}\delta_{l_{p}^{\prime}l_{p}}\delta_{j_{z}^{\prime}j_{z}%
},\label{rs1b}%
\end{equation}
where $d\mathbf{\omega=}d\cos\theta\ d\varphi.$ Then the normalization%
\begin{equation}
\int\Psi_{j^{\prime}l_{p}^{\prime}j_{z}^{\prime}}^{+}\left(  \epsilon
,\mathbf{\omega}\right)  \Psi_{jl_{p}j_{z}}\left(  \epsilon,\mathbf{\omega
}\right)  d^{3}\mathbf{p}=\delta_{j^{\prime}j}\delta_{l_{p}^{\prime}l_{p}%
}\delta_{j_{z}^{\prime}j_{z}}\label{rs1c}%
\end{equation}
implies the condition for the amplitude $\phi_{j},$
\begin{equation}
\int\phi_{j}^{\ast}\left(  \epsilon\right)  \phi_{j}\left(  \epsilon\right)
p^{2}dp=1.\label{rs1d}%
\end{equation}
In the next discussion it will be convenient also to use the alternative
representation, which differs in normalization,
\begin{equation}
a_{j}\left(  \epsilon\right)  =\frac{\phi_{j}\left(  \epsilon\right)  }%
{2\sqrt{\pi}};\qquad\int a_{j}^{\ast}\left(  \epsilon\right)  a_{j}\left(
\epsilon\right)  d^{3}\mathbf{p}=1.\label{rs1f}%
\end{equation}

\subsection{Angular moments of one-fermion states}

A few examples of the corresponding probability distribution%
\begin{equation}
P_{j,j_{z}}\left(  \mathbf{\omega}\right)  =\Phi_{jl_{p}j_{z}}^{+}\left(
\mathbf{\omega}\right)  \Phi_{jl_{p}j_{z}}\left(  \mathbf{\omega}\right)
;\ \int P_{j,j_{z}}\left(  \mathbf{\omega}\right)  d\mathbf{\omega
}=1,\label{rs2}%
\end{equation}
are given in Table \ref{tb1}. These distributions does not depend on the
parameters $\varphi$\ and $l_{p}=j\pm1/2$. \begin{table}[ptb]
\begin{center}%
\begin{tabular}
[c]{c|c|}%
$j,j_{z}$ & $P_{j,j_{z}}(\omega)$\\\hline
$\frac{1}{2},\frac{1}{2}$ & $1$\\
$\frac{3}{2},\frac{3}{2}$ & $\frac{3-3\cos2\theta}{4}$\\
$\frac{3}{2},\frac{1}{2}$ & $\frac{5+3\cos2\theta}{4}$\\
$\frac{5}{2},\frac{5}{2}$ & $\frac{45-60\cos2\theta+15\cos4\theta}{64}$%
\end{tabular}
\end{center}
\caption{The examples of the distributions (\ref{rs2}). The common factor
$1/4\pi$ is omitted.}%
\label{tb1}%
\end{table}The lowest value $j=1/2$ generates rotational symmetry of the
probability distribution, but for higher $j=3/2,5/2,...$ the distribution has
axial symmetry only. The states (\ref{rs1}) are not eigenstates of spin and
OAM; nevertheless, one can always calculate the mean values of corresponding
operators%
\begin{equation}
s_{z}=\frac{1}{2}\left(
\begin{array}
[c]{cc}%
\sigma_{z} & 0\\
0 & \sigma_{z}%
\end{array}
\right)  ,\ l_{z}=-i\left(  p_{x}\frac{\partial}{\partial p_{y}}-p_{y}%
\frac{\partial}{\partial p_{x}}\right)  .\label{rs3}%
\end{equation}
The related matrix elements are given by the relations \cite{prd13}:%
\begin{align}
\left\langle s_{z}\right\rangle _{j,j_{z}} &  =\int\Phi_{jl_{p}j_{z}}^{+}%
s_{z}\Phi_{jl_{p}j_{z}}d\mathbf{\omega}=\frac{1+\left(  2j+1\right)  \mu
}{4j\left(  j+1\right)  }j_{z},\label{rs4}\\
\left\langle l_{z}\right\rangle _{j,j_{z}} &  =\int\Phi_{jl_{p}j_{z}}^{+}%
l_{z}\Phi_{jl_{p}j_{z}}d\mathbf{\omega}\nonumber\\
&  =\left(  1-\frac{1+\left(  2j+1\right)  \mu}{4j\left(  j+1\right)
}\right)  j_{z},\nonumber
\end{align}
in which we have denoted
\begin{equation}
\mu=\pm\frac{m}{\epsilon},\label{rs4a}%
\end{equation}
where the sign $\left(  \pm\right)  $ corresponds to $l_{p}=j\mp1/2$. The
relations imply that in the nonrelativistic limit, when $\mu\simeq\pm1,$ we
get for signs $\left(  \pm\right)  $ \ correspondingly,%
\begin{equation}
\left\langle s_{z}\right\rangle _{j,j_{z}}=%
\genfrac{\{}{\}}{0pt}{}{\frac{j_{z}}{2j}}{\frac{-j_{z}}{2\left(  j+1\right)
}}%
,\ \left\langle l_{z}\right\rangle _{j,j_{z}}=%
\genfrac{\{}{\}}{0pt}{}{\left(  1-\frac{1}{2j}\right)  j_{z}}{\left(
1+\frac{1}{2\left(  j+1\right)  }\right)  j_{z}}%
\label{rs6}%
\end{equation}
and in the relativistic case, when $\mu\rightarrow0,$ we have%
\begin{equation}
\left\langle s_{z}\right\rangle _{j,j_{z}}=\frac{j_{z}}{4j\left(  j+1\right)
},\ \left\langle l_{z}\right\rangle _{j,j_{z}}=\left(  1-\frac{1}{4j\left(
j+1\right)  }\right)  j_{z}.\label{rs7}%
\end{equation}
The last two relations imply%
\begin{align}
\left\vert \left\langle s_{z}\right\rangle _{j,j_{z}}\right\vert  &  \leq
\frac{1}{4\left(  j+1\right)  }\leq\frac{1}{6},\label{rs8}\\
\frac{\left\vert \left\langle s_{z}\right\rangle _{j,j_{z}}\right\vert
}{\left\vert \left\langle l_{z}\right\rangle _{j,j_{z}}\right\vert } &
\leq\frac{1}{4j^{2}+4j-1}\leq\frac{1}{2}.\nonumber
\end{align}
For the complete wave function (\ref{rs1a}), the relations (\ref{rs4}) are
modified as
\begin{align}
\left\langle \left\langle s_{z}\right\rangle \right\rangle _{j,j_{z}} &
=\int\Psi_{_{jl_{p}j_{z}}}^{+}s_{z}\Psi_{jl_{p}j_{z}}d^{3}\mathbf{p}%
=\frac{1+\left(  2j+1\right)  \left\langle \mu_{j}\right\rangle }{4j\left(
j+1\right)  }j_{z},\label{rs8a}\\
\left\langle \left\langle l_{z}\right\rangle \right\rangle _{j,j_{z}} &
=\int\Psi_{_{jl_{p}j_{z}}}^{+}l_{z}\Psi_{jl_{p}j_{z}}d^{3}\mathbf{p}%
\nonumber\\
&  =\left(  1-\frac{1+\left(  2j+1\right)  \left\langle \mu_{j}\right\rangle
}{4j\left(  j+1\right)  }\right)  j_{z},\nonumber
\end{align}
where%
\begin{equation}
\left\langle \mu_{j}\right\rangle =\pm\int a_{j}^{\ast}\left(  \epsilon
\right)  a_{j}\left(  \epsilon\right)  \frac{m}{\epsilon}d^{3}\mathbf{p}%
,\qquad\left\vert \left\langle \mu_{j}\right\rangle \right\vert \leq
1.\label{rs8b}%
\end{equation}

\subsection{Many-fermion states}

The system of fermions (or arbitrary particles) generating the state with
quantum numbers $J,J_{z}$ can be represented by the combination of
one-particle states. For example the pair of states $j_{1},j_{2}$ can generate
the states%
\begin{gather}
\left\vert (j_{1},j_{2})J,J_{z}\right\rangle \label{rs9}\\
=\sum_{j_{z1}=-j_{1}}^{j_{1}}\sum_{j_{z2}=-j_{2}}^{j_{2}}\left\langle
j_{1},j_{z1},j_{2},j_{z2}\left\vert J,J_{z}\right.  \right\rangle \left\vert
j_{1},j_{z1}\right\rangle \left\vert j_{2},j_{z2}\right\rangle ;\nonumber\\
j_{z1}+j_{z2}=J_{z},\qquad\left\vert j_{1}-j_{2}\right\vert \leq J\leq
j_{1}+j_{2},\label{rs9a}%
\end{gather}
where $\left\langle j_{1},j_{z1},j_{2},j_{z2}\left\vert J,J_{z}\right.
\right\rangle $ are Clebsch-Gordan coefficients, which are nonzero if the
conditions (\ref{rs9a}) are satisfied. In this way one can repeat the
composition and obtain the many-particle eigenstates of resulting $J,J_{z}$%
\begin{align}
&  \left\vert (j_{1},j_{2},...j_{n})_{c}J,J_{z}\right\rangle \label{rs10}\\
&  =\sum_{j_{z1}=-j_{1}}^{j_{1}}\sum_{j_{z2}=-j_{2}}^{j_{2}}...\sum
_{j_{zn}=-j_{n}}^{j_{n}}c_{j}\left\vert j_{1},j_{z1}\right\rangle \left\vert
j_{2},j_{z2}\right\rangle ...\left\vert j_{n},j_{zn}\right\rangle ,\nonumber
\end{align}
where the coefficients $c_{j}$ are a product of the Clebsch-Gordan
coefficients%
\begin{align}
c_{j} &  =\left\langle j_{1},j_{z1},j_{2},j_{z2}\left\vert J_{3}%
,J_{3z}\right.  \right\rangle \left\langle J_{3},J_{z3},j_{3},j_{z3}\left\vert
J_{4},J_{z4}\right.  \right\rangle \label{rs10b}\\
&  ...\left\langle J_{n},J_{zn},j_{n},j_{zn}\left\vert J,J_{z}\right.
\right\rangle .\nonumber
\end{align}
Let us remark that the set $j_{1},j_{2},..j_{n}$ does not define the resulting
state unambiguously. The result depends on the pattern of \ their composition,
e.g.
\begin{align}
&  (((j_{1}\oplus j_{2})_{J_{1}}\oplus j_{3})_{J_{2}}\oplus j_{4}%
)_{J},\label{10d}\\
&  (((j_{1}\oplus j_{2})_{J_{1}}\oplus(j_{3}\oplus j_{4})_{J_{2}})_{J_{3}%
}\oplus j_{5})_{J},\nonumber
\end{align}
where $J_{k}$ represent intermediate AMs corresponding to the steps of
composition:%
\begin{equation}
j_{1}\oplus j_{2}=J_{1},\qquad J_{1}\oplus j_{3}=J_{2},\qquad J_{2}\oplus
j_{4}=J.\label{10e}%
\end{equation}
Each binary composition "$\oplus$"\ is defined by Eq. (\ref{rs9}). Different
composition patterns are in (\ref{rs10}) symbolically expressed by the
subscript $c$. Apparently, the number of patterns increases with $n$ very
rapidly; however, in a real scenario with an interaction one can expect their
probabilities will differ. The case $n=3$ will be illustrated in more detail below.

From now we discuss only the composed states with resulting $J=J_{z}=1/2$
($J_{z}=-1/2$ gives the equivalent results). The corresponding $n$-fermion
state ($n$ is odd)%
\begin{equation}
\Phi_{c,1/2,1/2}(\mathbf{\omega}_{1},\mathbf{\omega}_{2},..\mathbf{\omega}%
_{n})=\left\vert (j_{1},j_{2},...j_{n})_{c}1/2,1/2\right\rangle ,
\label{rs10a}%
\end{equation}
or alternatively%
\begin{align}
\Psi_{c,1/2,1/2}  &  =\phi_{j_{1}}\left(  \epsilon_{1}\right)  \phi_{j_{2}%
}\left(  \epsilon_{2}\right)  ..\phi_{j_{n}}\left(  \epsilon_{n}\right)
\label{rs10c}\\
&  \times\Phi_{c,1/2,1/2}(\mathbf{\omega}_{1},\mathbf{\omega}_{2}%
,..\mathbf{\omega}_{n})\nonumber
\end{align}
generate the $n$-dimensional angular distribution
\begin{equation}
P_{c}(\mathbf{\omega}_{1},\mathbf{\omega}_{2},..\mathbf{\omega}_{n}%
)=\Phi_{c,1/2,1/2}^{+}\Phi_{c,1/2,1/2}, \label{rs11}%
\end{equation}
from which the corresponding average one-fermion distributions are obtained as%
\begin{equation}
p_{c,k}(\mathbf{\omega}_{k})=\int P_{c}(\mathbf{\omega}_{1},\mathbf{\omega
}_{2},..\mathbf{\omega}_{n})\prod_{i\neq k}^{n}d\mathbf{\omega}_{i},
\label{rs11a}%
\end{equation}
which gives \cite{prd13}:%

\begin{equation}
p_{c,k}(\mathbf{\omega})=\frac{1}{4\pi}. \label{RS11G}%
\end{equation}
It follows that the distribution%

\begin{equation}
P_{c}(\mathbf{\omega})=\sum_{k=1}^{n}p_{c,k}(\mathbf{\omega})=\frac{n}{4\pi},
\label{rs16b}%
\end{equation}
which is generated by the state (\ref{rs10a}) has rotational symmetry similar
to the distribution $P_{1/2,1/2}$ generated by the one-fermion state in Table
\ref{tb1}. Therefore the angular probability distribution $P_{c}%
(\mathbf{\omega})$ related to the state $J=1/2$ has rotational symmetry
regardless of the number of involved particles. This rule suggests that e.g.
in a nucleus $J=1/2$, the probability distribution of nucleons, separately for
protons and neutrons, has in the momentum space rotational symmetry. Spherical
symmetry of probability distribution in the momentum space apparently implies
spherical symmetry in coordinate representation.

What can be said about the mean values of the spin and OAM contributions
\begin{gather}
\left\langle \mathbb{S}_{z}\right\rangle _{c,1/2,1/2}=\left\langle
s_{z1}+s_{z2}+...+s_{zn}\right\rangle _{c},\label{rs16}\\
\left\langle \mathbb{L}_{z}\right\rangle _{c,1/2,1/2}=\left\langle
l_{z1}+l_{z2}+...+l_{zn}\right\rangle _{c},\nonumber\\
\left\langle \mathbb{S}_{z}\right\rangle _{c,1/2,1/2}+\left\langle
\mathbb{L}_{z}\right\rangle _{c,1/2,1/2}=\frac{1}{2},\nonumber
\end{gather}
corresponding to the state (\ref{rs10a})? Now we will discuss this question in
more detail for the case $n=3$. 

\subsubsection{Three-fermion states}

There are three patterns for composition of\ the three AMs $j_{a},j_{b},j_{c}%
$:%
\begin{equation}
(\left(  j_{a}\oplus j_{b}\right)  _{J_{c}}\oplus j_{c})_{1/2};\qquad
abc=123,312,231.\label{ri1}%
\end{equation}
Corresponding states are
\begin{align}
&  \Phi_{c,1/2,1/2}(\mathbf{\omega}_{1},\mathbf{\omega}_{2},\mathbf{\omega
}_{3})\label{rs12}\\
&  =\sum_{j_{z1}=-j_{1}}^{j_{1}}\sum_{j_{z2}=-j_{2}}^{j_{2}}\sum
_{j_{z3}=-j_{3}}^{j_{3}}\left\langle j_{a},j_{za},j_{b},j_{zb}\left\vert
J_{c},J_{zc}\right.  \right\rangle \nonumber\\
&  \times\left\langle J_{c},J_{zc},j_{c},j_{zc}\left\vert 1/2,1/2\right.
\right\rangle \left\vert j_{1},j_{z1}\right\rangle \left\vert j_{2}%
,j_{z2}\right\rangle \left\vert j_{3},j_{z3}\right\rangle .\nonumber
\end{align}
The conditions (\ref{rs9a}) give at most two possibilities for the
intermediate values $J_{c},$ which must satisfy
\begin{equation}
J_{c}=j_{c}\pm1/2,\qquad\left\vert j_{a}-j_{b}\right\vert \leq J_{c}\leq
j_{a}+j_{b}.\label{rs13}%
\end{equation}
At the same time it holds%
\begin{equation}
j_{z1}+j_{z2}+j_{z3}=1/2,\qquad j_{za}+j_{zb}=J_{zc}.\label{rs15}%
\end{equation}
In this way two possible values $J_{c}$ in three patterns (\ref{ri1}) give six
possibilities to create the state (\ref{rs12}). Further, if we take into
account two possible values $l_{p}=j\pm1/2$ for each one-fermion state in
(\ref{rs12}) and defined by (\ref{rs1}), then in general the total number of
generated three-fermion states is $6\times2^{3}=48$. Due to orthogonality of
the terms in sum (\ref{rs12}) the three-fermion mean values (\ref{rs16}) are
calculated as
\begin{align}
&  \left\langle \mathbb{S}_{z}\right\rangle _{c,1/2,1/2}\label{rs17}\\
&  =\sum_{j_{z1}=-j_{1}}^{j_{1}}\sum_{j_{z2}=-j_{2}}^{j_{2}}\sum
_{j_{z3}=-j_{3}}^{j_{3}}\left\langle j_{a},j_{za},j_{b},j_{zb}\left\vert
J_{c},J_{zc}\right.  \right\rangle ^{2}\nonumber\\
&  \times\left\langle J_{c},J_{zc},j_{c},j_{zc}\left\vert 1/2,1/2\right.
\right\rangle ^{2}\left(  \left\langle s_{za}\right\rangle +\left\langle
s_{zb}\right\rangle +\left\langle s_{zc}\right\rangle \right)  \nonumber
\end{align}
and similarly for $\left\langle \mathbb{L}_{z}\right\rangle _{c,1/2,1/2}$.
Corresponding one-fermion values $\left\langle s_{z..}\right\rangle $ and
$\left\langle l_{z..}\right\rangle $ are given by the relations (\ref{rs4}).
The results for a set of input values $j_{1},j_{2},j_{3}$ and $l_{pk}%
=j_{k}-1/2$ are listed in Table \ref{tb2} and the results corresponding to
remaining sets $l_{pk}=j_{k}\pm1/2$ are similar and differ only in terms
proportional to $\tilde{\mu}$. Since%
\begin{gather}
\left\langle \mathbb{S}_{z}\right\rangle _{c,1/2,1/2}=-\left\langle
\mathbb{S}_{z}\right\rangle _{c,1/2,-1/2},\label{rs18}\\
\left\langle \mathbb{S}_{z}\right\rangle _{c,1/2,\pm1/2}+\left\langle
\mathbb{L}_{z}\right\rangle _{c,1/2,\pm1/2}=\pm1/2,\nonumber
\end{gather}
we present only $\left\langle \mathbb{S}_{z}\right\rangle _{c}\equiv
\left\langle \mathbb{S}_{z}\right\rangle _{c,1/2,1/2}$. \begin{table}[ptb]
\begin{center}%
\begin{tabular}
[c]{ccc|ccc|ccc}%
$j_{1}$ & $j_{2}$ & $j_{3}$ & $\left\langle S_{z}\right\rangle _{3}$ &
$\left\langle S_{z}\right\rangle _{2}$ & $\left\langle S_{z}\right\rangle
_{1}$ & $\left\langle S_{z}\right\rangle _{3}$ & $\left\langle S_{z}%
\right\rangle _{2}$ & $\left\langle S_{z}\right\rangle _{1}$\\\hline
$\frac{1}{2}$ & $\frac{1}{2}$ & $\frac{1}{2}$ & $\frac{1+2\tilde{\mu}}{6}$ &
$\frac{1+2\tilde{\mu}}{6}$ & $\frac{1+2\tilde{\mu}}{6}$ & $\frac{1+2\tilde
{\mu}}{6}$ & $\frac{1+2\tilde{\mu}}{6}$ & $\frac{1+2\tilde{\mu}}{6}$\\
$\frac{3}{2}$ & $\frac{1}{2}$ & $\frac{1}{2}$ & $\times$ & $\times$ &
$\frac{-1}{18}$ & $\frac{-1}{18}$ & $\frac{-1}{18}$ & $\times$\\
$\frac{3}{2}$ & $\frac{3}{2}$ & $\frac{1}{2}$ & $\frac{1+2\tilde{\mu}}{6}$ &
$\frac{1+3\tilde{\mu}}{18}$ & $\frac{1+3\tilde{\mu}}{18}$ & $\frac
{-1+6\tilde{\mu}}{90}$ & $\frac{3+7\tilde{\mu}}{30}$ & $\frac{3+7\tilde{\mu}%
}{30}$\\
$\frac{3}{2}$ & $\frac{3}{2}$ & $\frac{3}{2}$ & $\frac{1+4\tilde{\mu}}{30}$ &
$\frac{1+4\tilde{\mu}}{30}$ & $\frac{1+4\tilde{\mu}}{30}$ & $\frac
{1+4\tilde{\mu}}{30}$ & $\frac{1+4\tilde{\mu}}{30}$ & $\frac{1+4\tilde{\mu}%
}{30}$\\
$\frac{5}{2}$ & $\frac{3}{2}$ & $\frac{1}{2}$ & $\times$ & $\times$ &
$\frac{-5-4\tilde{\mu}}{90}$ & $\frac{-5-4\tilde{\mu}}{90}$ & $\frac
{-5-4\tilde{\mu}}{90}$ & $\times$\\
$\frac{5}{2}$ & $\frac{3}{2}$ & $\frac{3}{2}$ & $\frac{5+17\tilde{\mu}}{90}$ &
$\frac{5+17\tilde{\mu}}{90}$ & $\frac{-1+2\tilde{\mu}}{90}$ & $\frac
{-1+29\tilde{\mu}}{630}$ & $\frac{-1+29\tilde{\mu}}{630}$ & $\frac
{41+134\tilde{\mu}}{630}$\\
$\frac{5}{2}$ & $\frac{5}{2}$ & $\frac{1}{2}$ & $\frac{1+2\tilde{\mu}}{6}$ &
$\frac{13+38\tilde{\mu}}{270}$ & $\frac{13+38\tilde{\mu}}{270}$ &
$\frac{-23+2\tilde{\mu}}{630}$ & $\frac{31+74\tilde{\mu}}{378}$ &
$\frac{31+74\tilde{\mu}}{378}$\\
$\frac{5}{2}$ & $\frac{5}{2}$ & $\frac{3}{2}$ & $\frac{29+104\tilde{\mu}}%
{630}$ & $\frac{23+152\tilde{\mu}}{1890}$ & $\frac{23+152\tilde{\mu}}{1890}$ &
$\frac{-1+8\tilde{\mu}}{210}$ & $\frac{55+232\tilde{\mu}}{1890}$ &
$\frac{55+232\tilde{\mu}}{1890}$\\
$\frac{5}{2}$ & $\frac{5}{2}$ & $\frac{5}{2}$ & $\frac{1+6\tilde{\mu}}{70}$ &
$\frac{1+6\tilde{\mu}}{70}$ & $\frac{1+6\tilde{\mu}}{70}$ & $\frac
{1+6\tilde{\mu}}{70}$ & $\frac{1+6\tilde{\mu}}{70}$ & $\frac{1+6\tilde{\mu}%
}{70}$%
\end{tabular}
\end{center}
\caption{Mean values $\left\langle \mathbb{S}_{z}\right\rangle _{c}$ of
three-fermion states $\left\vert (j_{1},j_{2},j_{3},J_{c})1/2,1/2\right\rangle
$ with $J_{c}=j_{c}-1/2$ and $J_{c}=j_{c}+1/2$ (columns 4,5,6 and 7,8,9;
$c=3,2,1$) [see the first relation (\ref{rs13}) and (\ref{rs17})]. The symbol
$\times$ denotes configuration for which the second condition (\ref{rs13}) is
not satisfied.}%
\label{tb2}%
\end{table}The meaning of the parameter $\tilde{\mu}$ is as follows:

(1) If one assumes the same parameter $\mu\ $(\ref{rs4a}) for the three
fermions in the state (\ref{rs12}), then $\tilde{\mu}=\mu.$

(2) In a general case, the complete wave function%
\begin{align}
\Psi_{c,1/2,1/2}  &  =\phi_{j_{1}}\left(  \epsilon_{1}\right)  \phi_{j_{2}%
}\left(  \epsilon_{2}\right)  \phi_{j_{3}}\left(  \epsilon_{3}\right)
\label{rs18a}\\
&  \times\Phi_{c,1/2,1/2}(\mathbf{\omega}_{1},\mathbf{\omega}_{2}%
,\mathbf{\omega}_{3})\nonumber
\end{align}
gives instead of (\ref{rs8b}) a more complicated expression \cite{prd13}%
\begin{equation}
\tilde{\mu}=f_{c}\left(  \left\langle \mu_{1}\right\rangle ,\left\langle
\mu_{2}\right\rangle ,\left\langle \mu_{3}\right\rangle ,j_{1},j_{2}%
,j_{3}\right)  , \label{rs18b}%
\end{equation}
where the parameters $\left\langle \mu_{i}\right\rangle $\ are defined by Eq.
(\ref{rs8b}). The expression is simplified for $\left\langle \mu
_{1}\right\rangle =\left\langle \mu_{2}\right\rangle =\left\langle \mu
_{3}\right\rangle =\left\langle \mu\right\rangle $,%
\begin{equation}
f_{c}\left(  \left\langle \mu\right\rangle ,\left\langle \mu\right\rangle
,\left\langle \mu\right\rangle ,j_{1},j_{2},j_{3}\right)  =\left\langle
\mu\right\rangle . \label{rs18c}%
\end{equation}
Obviously the many-fermion system with $J=J_{z}=1/2$ can be treated as a
composed particle of the spin $1/2$. This spin is generated by the spins and
OAMs of the involved fermions. The relative weights of the spin and OAM
contributions vary depending not only on the intrinsic values $j_{1}%
,j_{2},j_{3}$ and the pattern of composition, but also on the mass-motion
parameter $\tilde{\mu}$. The data in the table suggest that for any
configuration in the relativistic limit $\tilde{\mu}\rightarrow0,$ we have%
\begin{equation}
\left\vert \left\langle \mathbb{S}_{z}\right\rangle \right\vert \leq\frac
{1}{6} \label{rs19}%
\end{equation}
similar to the case of the one-fermion states (\ref{rs8}).

The table illustrates a complexity of the AM composition even for only three
fermions. Is there a simple rule like (\ref{rs19}) for $n>3$? First, let us
consider the composition%
\begin{equation}
\Psi_{c,1/2,1/2}=\left\vert (j_{1},j_{2},...j_{n})_{c}1/2,1/2\right\rangle ,
\label{rs21}%
\end{equation}
where all one-fermion AMs are the same, $j_{i\text{ }}=j$ (like the rows 1,4,9
in the table). The corresponding spin reads%
\begin{equation}
\left\langle \mathbb{S}_{z}\right\rangle =\frac{1+\left(  2j+1\right)
\tilde{\mu}}{8j\left(  j+1\right)  } \label{RS23}%
\end{equation}
regardless of $n$ and details of composition. The proof of this relation is
given in \cite{prd13}. Apparently for $\tilde{\mu}\rightarrow0,$%
\ the\ relation (\ref{rs19}) is again satisfied. The situation with the
composition of different AMs is getting much more complex for increasing $n$.
However, an average value of the spin over all possible composition patterns
of the state $\left\vert (j_{1},j_{2},...j_{n})_{c}1/2,1/2\right\rangle $
appears to safely satisfy (\ref{rs19}). This is the case when there is no
(e.g., dynamical) preference among various composition patterns.

Let us illustrate a possible role of the composition patterns by the simple
example $j_{1},j_{2},j_{3}=1/2$. Equation (\ref{rs12}) gives the three\ states
corresponding to $J_{c}=1$,%
\begin{gather}
\Psi_{abc,1/2,1/2}=\frac{\phi_{abc}}{\sqrt{6}}\left(  \left\vert
-1/2,1/2,1/2\right\rangle \right. \label{ri2}\\
+\left.  \left\vert 1/2,-1/2,1/2\right\rangle -2\left\vert
1/2,1/2,-1/2\right\rangle \right)  ,\nonumber
\end{gather}
where%
\begin{equation}
\phi_{abc}=\phi_{a}\left(  \epsilon_{a}\right)  \phi_{b}\left(  \epsilon
_{b}\right)  \phi_{c}\left(  \epsilon_{c}\right)  . \label{ri3}%
\end{equation}
The indices $abc$ define the composition in accordance with (\ref{ri1}), and
AM states are defined correspondingly, $\left\vert j_{za},j_{zb}%
,j_{zc}\right\rangle $. The other three states correspond to $J_{c}=0$,
\begin{equation}
\Psi_{abc,1/2,1/2}=\frac{\phi_{abc}}{\sqrt{2}}\left(  \left\vert
1/2,-1/2,1/2\right\rangle -\left\vert -1/2,1/2,1/2\right\rangle \right)  .
\label{ri4}%
\end{equation}
The nonrelativistic proton SU(6) wave function in a standard notation reads:%
\begin{gather}
\left\vert p\uparrow\right\rangle =\frac{1}{\sqrt{2}}\left\{  \frac{1}%
{\sqrt{6}}\left\vert duu+udu-2uud\right\rangle \right. \label{ri5}\\
\times\frac{1}{\sqrt{6}}\left\vert \downarrow\uparrow\uparrow+\uparrow
\downarrow\uparrow-2\uparrow\uparrow\downarrow\right\rangle \nonumber\\
+\left.  \frac{1}{\sqrt{2}}\left\vert duu-udu\right\rangle \frac{1}{\sqrt{2}%
}\left\vert \downarrow\uparrow\uparrow-\uparrow\downarrow\uparrow\right\rangle
\right\}  .\nonumber
\end{gather}
The comparison (\ref{ri2})$-$(\ref{ri4}) with (\ref{ri5}) suggests that the
SU(6) wave function after substitution%
\[
\phi_{a}\left(  \epsilon_{a}\right)  =u_{1},\qquad\phi_{b}\left(  \epsilon
_{b}\right)  =u_{2,}\qquad\phi_{c}\left(  \epsilon_{c}\right)  =d
\]
can be obtained as the superposition of wave functions generated by the AM
compositions%
\begin{gather}
(\left(  u_{1}\oplus u_{2}\right)  _{J}\oplus d)_{1/2},\qquad(\left(  d\oplus
u_{1}\right)  _{J}\oplus u_{2})_{1/2},\label{ri6}\\
(\left(  u_{2}\oplus d\right)  _{J}\oplus u_{1})_{1/2}\nonumber
\end{gather}
for $J=1,2$.

\section{Conclusion}

Our study was focused on the many-fermion system carrying spin $J=1/2$,
however the relation (\ref{rs19}) can be generalized for arbitrary spin $J$%
\begin{equation}
\left\vert \left\langle \mathbb{S}_{z}\right\rangle \right\vert \leq\frac
{J}{3}\label{ri7}%
\end{equation}
provided that:

(1) the intrinsic motion of the fermions inside the system (composite
particle) is relativistic $\left(  \tilde{\mu}\rightarrow0\right)  $,

(2) mean value $\left\langle \mathbb{S}_{z}\right\rangle $ include an
averaging over possible composition patterns (if the number of fermions
$n\geq3$)

The ratio $\tilde{\mu}=\left\langle m/\epsilon\right\rangle $ is of key
importance, since it controls a "contraction" of the spin component
(\ref{ri7}), which is compensated by the OAM. It is a pure effect of
relativistic quantum mechanic. The obtained results for $J=1/2$ have been
applied to the description of the proton spin structure in terms of the
structure functions $g_{1}$ and \ $g_{2}$ in Ref. \cite{prd13}, where we have
suggested the proton studied at polarized deep inelastic scattering is an
ideal instrument for the study of this relativistic effect.

This work was supported by the project LG130131 of Ministry of Education,
Youth and Sports of the Czech Republic.

\end{document}